\documentclass[twocolumn,showpacs,preprintnumbers,amsmath,amssymb]{revtex4}

\usepackage{graphicx}
\usepackage{dcolumn}
\usepackage{bm}

\begin{document}


\title{Nonlinear Rabi Oscillations of Excitons in Dense Quantum Dot System}

\author{Y. Mitsumori}
 \email{mitumori@crl.go.jp}

\author{A. Hasegawa}%
\author{M. Sasaki}%
\affiliation{%
Communications Research Laboratory, 4-2-1 Nukuikita, Koganei-shi, Tokyo 184-8795, Japan.}%

\author{H. Maruki}
\author{F. Minami}

\affiliation{
Department of Physics, Tokyo Institute of Technology, 2-12-1 Oh-okayama, Meguro-ku, Tokyo 152-8551, Japan.
}%

\date{\today}

\begin{abstract}
Nonlinear optical response and optical coherence of excitons in semiconductor dense quantum dots in GaAs single quantum wells has been studied by using photon echo techniques. At low temperatures, the optical coherence is estimated to be 2 ns from the decay curve in the photon echo signal. Rabi oscillations are also observed. Unlike the results for a single quantum dot in the literature to date, the oscillation shows nonlinear behavior affected strongly by the pre-excited carrier density. The results can be explained by the theory of the local field correction originating from exciton-exciton interaction between the inter dots.\end{abstract}

\maketitle
\indent During the last decade, various results of optical coherent manipulation of excitons in semiconductors have been reported, such as: indirect \cite{1,2} and direct \cite{3} observations of excitonic Rabi oscillations in higher dimensional semiconductor structures, manipulations of a one-qubit rotation of an exciton in a single quantum dot \cite{4, 5, 6, 7,8}, direct observations of excitonic Rabi oscillations in quantum dot ensemble \cite{9}, and entanglement manipulation in quantum dot \cite{10} and dot molecules \cite{11}. These results open great opportunities for quantum information processing based on semiconductor devices which have potential advantages, including the ease of integration and the use of matured industrial technologies. In particular, excitons make a clear distinction from other two level systems for qubits in the sense that they can effectively interact with photons, that is the signal carrier wave to build a quantum communication network, by their large dipole moment \cite{6, 7, 8, 9, 12}. The main obstacle is then to overcome the fast dephasing time of excitons due to the phonon scattering and the strong coupling with various charged excitations in solids via Coulomb interaction. There is, however, a dilemma such that while an exciton-exciton interaction is a key process for quantum gating, strong interaction usually causes fast dephasing. The coupling strength and the dephasing time also strongly depend on quantum confinement structures, such as quantum well, wire, dot and their shapes. To find the structures with long coherence time and strong coupling at the same time in semiconductors is an important problem for progress in both nanoscale semiconductor physics and development of quantum information technology.\\
\indent We report the observation of excitonic nonlinear Rabi oscillations which exhibit the effect of strong exciton-exciton interaction and at the same time ultra-long coherence (2 ns) in a quantum dot system in GaAs/Al$_{0.3}$Ga$_{0.7}$As single quantum well structure. The observed dephasing time in our sample is much longer than the values reported in the literature \cite{13, 14, 15, 16, 17},  in spite of the underlying strong interaction among excitons.\\ 
\indent Our samples are GaAs/Al$_{0.3}$Ga$_{0.7}$As single quantum wells grown on (001)-GaAs substrates by molecular-beam epitaxy (MBE) whose well width are 110 {\AA} and 90 {\AA}. Since fluctuations exist in the well width with one or two monolayers in both samples, the heavy hole exciton splits into three states. In case of 110 {\AA} well width sample, one is situated at the small luminescence peak around 1.540 eV in the macroscopic photo-luminescence (PL) spectrum, which comes from the two monolayer height region, and the other are the main luminescence peak which is in agreement with the lower peak of the macroscopic photo-luminescence excitation (PLE) spectrum (one monolayer height region) and the higher peak in the PLE spectrum, as shown in Fig. 1. The micro-PL spectrum shows many small peaks, corresponding to the individual excitonic transitions in each dot formed by the potential fluctuations due to the variations of the well width. This fact indicates that a dense quantum dot system is realized in the quantum well. The 90 {\AA} well width sample also shows the same properties as the 110 {\AA} well width sample.\\
\indent The excitation light source is a cw mode-locked Ti:Sapphire laser pumped by a frequency-doubled Nd:YVO$_{4}$ laser. The temporal duration of the generated pulse, $\tau_{L}$, is about 100 fs with the spectral width of $\sim$10 meV. The center of the excitation wavelength is resonant with the main luminescence peak of the single quantum well. The experiment is based on the photon echo technique using three pulses linearly polarized along the same direction. The signal intensity in the backward diffraction geometry in the direction {\bf $k_{3}+k_{2}-k_{1}$} is detected as a function of the time delay, $\tau_{1}$, between the first and second pulses. All measurements are performed at 7 K.\\
\begin{figure}
\includegraphics[width=0.6\linewidth]{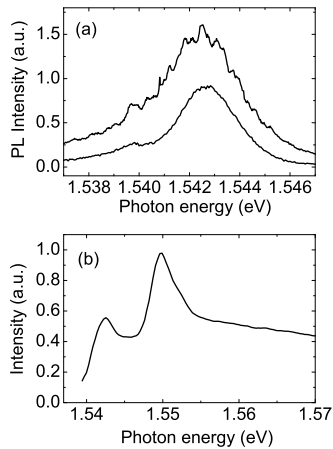}
\caption{(a) Macroscopic photo-luminescence (PL) spectrum (lower curve) together with micro-PL spectrum (upper curve) with the spatial resolution of 0.5 $\mu$m for the 110 {\AA} well width sample obtained at 7K. (b) Macroscopic photo-luminescence excitation (PLE) spectrum for the 110 {\AA} well width sample detected at 1.539 eV at 7K.}
\end{figure}
\begin{figure}
\includegraphics[width=0.7\linewidth]{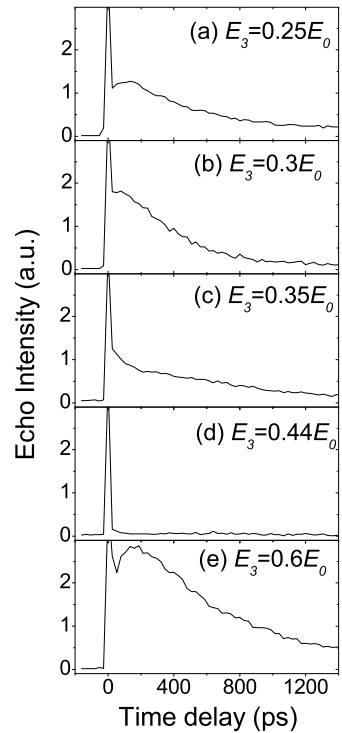}
\caption{Photon echo intensity for the 110 {\AA} well width sample detected as a function of the time delay between the first and second pulses for the various input power of the third pulse at 7K. The time delay between the second and third pulse is set to 0.3 ps. The input electric fields of the first and second pulses are fixed at $E_{1}$=0.3$E_{0}$ and $E_{2}$=0.7$E_{0}$, respectively. $E_{0}$ represents the maximum input electric field density of $\approx$ 2$\times$10$^{2}$ N/C/cm$^{2}$.}
\end{figure}
\begin{figure}
\includegraphics[width=0.7\linewidth]{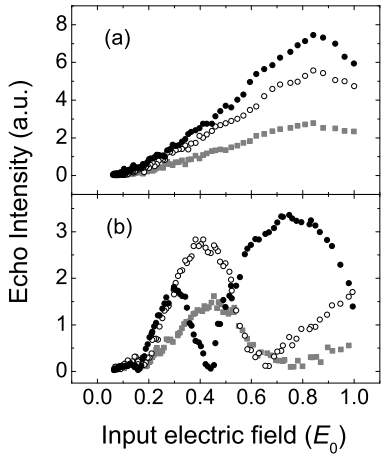}
\caption{Dependence of the photon echo intensity for the 110 {\AA} well width sample on the input electric field of the first pulse $E_{1}$ (a) and $E_{3}$ (b) at 7 K. The time delay between the first and second pulses is fixed at 100 ps and the second and third pulses at 0.3 ps. The input electric fields of the second pulse are set to $E_{2}$=0.1$E_{0}$  (gray squares), 0.3$E_{0}$ (open circles) and 0.7$E_{0}$ (closed circles), respectively. For (a), the third pulse is fixed at 0.3$E_{0}$ and for (b), the first pulse is set to $E_{1}$=0.3$E_{0}$. $E_{0}$ represents the maximum input electric field density of $\approx$ 2$\times$10$^{2}$ N/C/cm$^{2}$.}
\end{figure}
\indent Figure 2 shows the dependence of the decay profiles of the photon echo signal for the 110 {\AA} well width sample as a function of the time delay $\tau_{1}$ on the various input electric fields of the third pulse, $E_{3}$, when the input electric field of the first and second pulses are fixed at $E_{1}=0.3E_{0}$ and $E_{2}=0.7E_{0}$, respectively. Here, $E_{0}$ represents the maximum input electric field density of $\approx$ 2$\times$10$^{2}$ N/C/cm$^{2}$. The time delay $\tau_{2}$ between the second and third pulses is set to 0.3 ps so that the excitation pulse order can be defined exactly. A dominant signal at zero time delay in each scan is due to the photon echo from the GaAs substrate and/or a non-resonant two-photon transition to the GaAs substrate and is unrelated to the quantum dots \cite{18,19}. After these dominant peaks, the echo signals exhibit unusual decay profiles up to 200 ps, i.e. slight increase in (a), (b), and (e) while rapid decease in (c), which differ from conventional exponential decay. This will be discussed later again. For longer delays, we can observe a slowly decaying signal with a time constant of $\sim$500 ps. This signal is due to the dephasing of the exciton coherence in the quantum dots. Assuming the presence of inhomogeneous broadening, the coherence time can be calculated to be $\sim$2 ns \cite{20}, which corresponds to the homogeneous linewidth $\sim$0.65 $\mu$eV. The signal intensity and the decay profiles show the strong dependence on the input electric field of the third pulse. The signal intensity increases with  $E_{3}$ between 0.25$E_{0}$ and 0.3$E_{0}$. As $E_{3}$ further increases, the signal turns to decrease, completely disappears at $E_{3}$=0.44$E_{0}$, and then appears again. This behavior strongly suggests the Rabi oscillations expected for the coherent population oscillations of a two-level system with increasing pumping electric field intensity.\\
\indent Figure 3 shows the echo intensity for the 110 {\AA} well width sample as a function of the input electric field,  $E_{1}$ (a) or $E_{3}$ (b), at the fixed time delay $\tau_{1}$=100 ps and $\tau_{2}$=0.3 ps. In Fig. (a), the gray squares, the open and closed circles correspond to the cases of $E_{2}$=0.1$E_{0}$, $E_{2}$=0.3$E_{0}$ and $E_{2}$=0.7$E_{0}$, respectively, when $E_{3}$ is fixed at 0.3$E_{0}$. These show similar behavior on $E_{1}$, peaked around $E_{1}$=0.82$E_{0}$. In Fig. (b), the three cases $E_{2}$=0.1$E_{0}$ (gray squares) $E_{2}$=0.3$E_{0}$ (open circles), and $E_{2}$=0.7$E_{0}$ (closed circles) are compared, when fixing $E_{1}$=0.3$E_{0}$. It is clearly seen that the behavior on $E_{3}$ strongly depends on the second pulse intensity. The stronger the second pulse intensity is, the shorter the oscillation. The period and the amplitude in each scan are also found to be increasing with $E_{3}$. The observed nonlinear oscillation behavior is much different from that of an ideal two-level system \cite{21}, where the amplitude and period of the Rabi oscillation cannot change for the input field intensity, and from the results reported in the previous works for a single dot \cite{6,8} and a well isolated dot system \cite{9}. We have performed the same experiment on a dense quantum dot system in the 90 {\AA} well sample. The coherence time shows ultralong properties (2 ns) that are the same as the 110 {\AA} well width sample and the nonlinear Rabi oscillations are also observed.\\ 
\indent One plausible origin of the nonlinear oscillation behavior is the exciton-exciton interaction. In fact, the Coulomb renormalization of Rabi frequency has already been observed in semiconductor multiple quantum wells \cite{3} and in a well isolated dot system, the Rabi oscillations under the high excitation density is also modified by the biexciton effect, which corresponds to the exciton-exciton interaction (correlation) in intra dot \cite{9}. We can first rule out the biexciton effect since the oscillatory structure is still observable under the co-circularly polarized excitation (bi-exciton transition forbidden configuration). Further, the excitation density estimated from the present input electric field and the dipole moment $\mu$=15 Debye, calculated from the dephasing time of 2 ns, stays in the weak regime. This effect, therefore, is most likely attributed to the local field correction originating from exciton-exciton (dipole-dipole) interaction between the inter dots. Actually, according to the theoretical work \cite{27}, the local field effect modifies the Rabi oscillations. Using the rotating wave approximation, the self-consistent Maxwell-Bloch equations with local field correction in the resonant excitation can be written as \cite{22, 23, 24}\\
\begin{subequations}
\begin{eqnarray}
   \frac{\partial u}{ \partial t} & = & -\varepsilon w v
\\
   \frac{\partial v}{ \partial t} & = & \varepsilon w u + \Omega w
\\
   \frac{\partial w}{ \partial t} & = & -\Omega v
\end{eqnarray}
\end{subequations}
where $u$ ($v$ ) and $w$ represent the real (imaginary) part of the polarization and the population inversion, respectively. $\Omega$ denotes the Rabi frequency defined by $\mu$$E$/$\hbar$ using a dipole moment $\mu$ and an input electric field $E$. $\varepsilon$ represents, in the unit of frequency, the coupling constant of the near-dipole-dipole local field parameter arising from the local field  {\bf $E_{L}$}={\bf $E$}+{\bf $E$}$_{{\rm near}}$$-${\bf $E$}$_{P}$. {\bf $E$}$_{{\rm near}}$ represents the dipole field at a point {\bf $r$} generated by a microscopic polarization {\bf $p$} at a point {\bf $r$}$_{i}$, and is given by {\bf $E$}$_{{\rm near}}$=$\sum_{i}$\{(3{\bf $p$}$\cdot${\bf $n$}){\bf $n$}$-${\bf $p$}\}/(4$\pi$$\epsilon_{b}$$|${\bf $r$}$-${\bf $r$}$_{i}$$|$$^3$) \cite{25}. $\epsilon_{b}$ and {\bf $n$} are the dielectric constant of the background and an unit vector in the direction {\bf $r-r$}$_{i}$, respectively. {\bf $E_{P}$} is the average dipole field $-$$N${\bf $p$}/3$\epsilon_{b}$ using a dipole density $N$ \cite{25}. In the case where the isotropic three-dimensional dipole system, {\bf $E$}$_{{\rm near}}$  vanishes due to the symmetry \cite{25} and the coupling constant is given by $\varepsilon$=$\mu^2$$N$/3$\epsilon_{b}$$\hbar$ \cite{23,24}. For low dimensional systems, such as the present case, the contribution of the dipole field generated by microscopic polarization to the local field becomes dominant. Assuming a dipole system on a simple isotropic two-dimensional square lattice with a lattice spacing $d$, the coupling constant can be calculated and approximately given by $\varepsilon$=$\mu^{2}$/4$\epsilon_{b}$$\hbar$$d^{3}$. The ratio of the coupling constant and the Rabi frequency $\varepsilon$/$\Omega$ in the present case is estimated to be $>$0.1 from the average input electric field 0.3$E_{0}$, the dipole moment $\mu$=15 Debye and the average dot spacing $<$14 nm. The average dot spacing is estimated from the two-dimensional dot density of our sample $>$5$\times$10$^{11}$/cm$^{2}$ determined from the spectrum and spatial resolution limit of our micro-PL system. Thus the coupling constant of the local field correction is not negligible order compared with $\Omega$. Exciton dynamics in the present case may be described by the theory of the local field correction.\\
\indent The solution of Eq. (1) without the external electric field, that is the Rabi frequency, is found as $P(t)$ = $u$+$iv$ $\propto$ ${\rm exp}(-i \varepsilon w t)$. Thus the local field correction takes part in the optical response, the polarization rotates in the phase plane ($uv$-plane) with a frequency of $-$$\varepsilon w$. Assuming that polarization rotates slowly compared with the temporal pulse width, the effective Rabi frequency may be approximately written as\\
\begin{equation}
   \Omega_{{\rm eff}}=\sqrt{\Omega (\Omega + \varepsilon u_{0})}
\end{equation}
where $u_{0}$ represents the real part of the polarization just before the optical excitation. Thus the effective Rabi frequency increases with the real part of the pre-excited polarization, which is the positive value proportional to the square root of the pre-excited carrier density under the weak excitation regime. On the other hand, by applying the two-level model relating to the sinusoidal signal behavior \cite{26}, the photon echo intensity may be described by  $I_{{\rm echo}} \propto |{\rm sin} \theta_{{\rm eff}3} \times {\rm sin} \theta_{{\rm eff}2} \times {\rm sin} \theta_{{\rm eff}1}|^{2}$, using the effective pulse area of $\theta_{{\rm eff}i}$ = $\Omega_{{\rm eff}i}  \tau_{L}$ when a square pulse is assumed. This qualitatively explains the difference between Fig. 3 (a) and (b), namely, the echo intensity is insensitive to the first pulse where there exists no pre-excited carrier, while it is strongly dependent on the third pulse, and the oscillation period becomes shorter with increase in the second pulse field $E_{2}$. This pulse order effect can not be explained by the detuning term, $\Delta$, in an ideal two level system which gives the effective Rabi frequency $\Omega_{\rm eff}=\sqrt{\Omega^{2} + \Delta^{2}}$ insensitive to the pre-excited carrier density.\\
\indent Next, we focus on the nonlinear Rabi oscillations in Fig. 3 (b). The first terms in Eqs. (1a) and (1b) correspond to the detuning in an ideal two-level system, where the large (small) detuning causes the short (long) period and small (large) amplitude of the Rabi oscillations \cite{21}. In our case, the detuning varies with the population inversion, that is the excited carrier density. Under the weak (high) excitation range of the third pulse, the small (large) population inversion $w\approx -1$ ($w\approx 0$) leads to the large (small) detuning. The amplitude and period of the observed Rabi oscillations, therefore, increase with the input electric field of the third pulse.\\
\indent The unusual decay profiles up to 200 ps time delay as seen in Fig. 2 can also be explained in terms of the local field correction. The effective pulse area $\theta_{{\rm eff}}$, as mentioned before, depends on the real part of the pre-excited polarization which roughly evolves $u_{0}\propto {\rm sin}(- \varepsilon w_{1} \tau_{1})$ using the time delay $\tau_{1}$ and the population inversion $w_{1}$ after the first pulse. Under weak excitation of the first pulse, the population inversion takes a negative value, which results in the increase in the effective pulse area with the time delay $\tau_{1}$. In the case where $n \pi < \theta_{{\rm eff}3} < (n + \frac{1}{2} ) \pi$  ($n$ is integer) corresponding to Fig. 2 (a), (b), and (e), the larger effective pulse area of the third pulse for the longer time delay excites the system such that the echo intensity becomes stronger, leading to a slight increase in the echo intensity. On the other hand, in the region where $(n - \frac{1}{2} ) \pi< \theta_{{\rm eff}3} < n \pi$  corresponding to Fig. 2(c), the larger third pulse weakens the echo intensity, causing more rapid decrease than the exponential decay.\\
\indent Finally the problem of why the long coherence up to 2ns is possible even under the strong exciton-exciton interaction is still open. This problem may be essential for quantum information technology as well as fundamental semiconductor physics. It must be an important future work to measure coherence characteristics and coupling strengths for various dot systems whose dot spacing, shapes and arrangements are artificially controlled. Besides this further theoretical analysis on dephasing process including the strong dipole-dipole interaction is also necessary.\\
\indent We have shown the photon echo data in the dense quantum dot system in the single quantum wells. We found that the excitonic coherence created by laser light maintains 2 ns. The nonlinear Rabi oscillations, which strongly depend on the input pulse configuration and the pre-excited carrier density, are also observed in this system. The nonlinear behavior can be well explained by the theory of the local field correction originating from the exciton-exciton interaction between the inter dots.\\
\indent F. M. acknowledges for the partial support from a Grant-in-Aid for Scientific Research and Special Coordination Funds for Promoting Science and Technology from the STA (Science and Technology Agency) of Japan.

\newpage

\end{document}